\documentstyle[12pt,titlepage]{article}
\begin{document}
\begin{titlepage}
\parindent 0pt
{\Large\bf
Critical activity of the hard-sphere lattice gas
on the body-centred cubic lattice
}

\bigskip

{\large\bf Atsushi Yamagata}

{\it
Department of Physics, Tokyo Metropolitan University,
Minami-ohsawa 1-1, Hachioji-shi, Tokyo 192-03, Japan
}

\bigskip

\begin{description}
\item[Running title]
Critical activity of hard-sphere lattice gas

\item[Keywords]
Hard-sphere lattice gas, Critical phenomena, Monte Carlo method,
Antiferromagnetic Ising model, Blume-Emery-Griffiths model

\item[PACS classification codes]
02.70.Lq, 05.50.+q, 64.60.Cn, 68.35.Rh, 75.10.Hk
\end{description}

\bigskip

{\bf Abstract}

This is the first Monte Carlo study of the hard-sphere lattice gas
with nearest neighbour exclusion on the body-centred cubic lattice.
We estimate the critical activity to be $0.7223 \pm 0.0003$.
This result confirms that there is a re-entrant phase transition
of an antiferromagnetic Ising model in an external field and
a Blume-Emery-Griffiths model on the body-centred cubic lattice.
\end{titlepage}

\section{Introduction}
\label{sec:intro}
For phase transitions and critical phenomena of a hard-sphere
lattice gas~\cite{Domb58,Temperley59,Burley60}
while many authors have investigated two-dimensional systems
\cite{GauntFisher65,Runnels65,RunnelsCombs66,ReeChesnut66,Gaunt67,%
Baxteretal80,Baxter80,Racz80,WoodGoldfinch80,BaxterPearce82,%
Meirovitch83,BloteWu90,KamieniarzBlote93},
there are only a few studies in three dimensions
\cite{Gaunt67,Yamagata95a,Yamagata95b}.
One of the reason is that series expansions or transfer matrix methods
have been applied to the systems mainly.
We have obtained the critical activity and the critical exponents of
the hard-sphere lattice gas on the simple cubic lattice by using a
Monte Carlo method and a finite-size scaling
\cite{Yamagata95a,Yamagata95b}.
In this paper we perform Monte Carlo simulations on the body-centred
cubic lattice for the first time and estimate the critical activity.

Consider a hard-sphere lattice gas whose atoms occupy sites of
a lattice and interact with infinite repulsion on a site and of
nearest neighbour pairs.
The grand partition function is
\begin{equation}
{\mit \Xi_{V}(z)}
=
\sum_{N} z^{N}\,Z_{V}(N),
\label{eqn:gpf}
\end{equation}
where $z$ is an activity and
$Z_{V}(N)$ is the number of configurations in which
there are $N$ atoms in the lattice of $V$ sites.
At $z=+\infty$ a configuration of the ground state is that
the atoms occupy all the sites of one sublattice and
the other is vacant.
There is no atom at $z=0$.
A continuous phase transition occurs at a critical activity.

The hard-sphere lattice gas relates to an antiferromagnetic Ising
model in an external field
\cite{Temperley59,Baxteretal80,Racz80}.
The critical curve of the latter behaves as
\[
H=H_{\rm c} + a^{\ast} k_{\rm B} T
\]
near $T=0$, where $T$ is the temperature;
$H$ is the external field;
$k_{\rm B}$ is Boltzmann's constant;
$H_{\rm c}$ is the critical field at $T=0$.
In the present paper we shall consider only the case $H_{\rm c}>0$.
When $a^{\ast}$ is positive,
a re-entrant phase transition occurs.
The system is in the paramagnetic, the antiferromagnetic ordered, and
the paramagnetic phase as $T$ is decreased when $H$ is fixed slightly
above $H_{\rm c}$.
The slope, $a^{\ast}$, is given by the critical activity, $z_{\rm c}$,
\[
a^{\ast}=-\frac{1}{2}\ln z_{\rm c}.
\]
On the body-centred cubic lattice Gaunt
\cite{Gaunt67}
estimated $z_{\rm c}$ to be 0.77(5) by using series expansions of
the hard-sphere lattice gas and Landau
\cite{Landau77}
obtained $a^{\ast}=0.160(16)$, i.e., $z_{\rm c}=0.726(23)$ from
Monte Carlo simulations of the antiferromagnetic Ising model.
These results consist each other within errors.
There is a re-entrant phase transition since $a^{\ast}$ is positive.

In the next section we define physical quantities measured.
In section~\ref{sec:mcr} we present Monte Carlo results.
A summary is given in section~\ref{sec:sum}.

\section{Monte Carlo simulations}
\label{sec:mcs}
We use the Metropolis Monte Carlo technique
\cite{Binder79,BinderStauffer87}
to simulate the hard-sphere lattice gas~(\ref{eqn:gpf})
on the body-centred cubic lattice
of $V$ sites, where $V = 2 \times L \times L \times L$
($L$ = $2 \times n$, $n$ = 2, 3, $\ldots$ , 12),
under fully periodic boundary conditions.
A body-centred cubic lattice is made up two $L \times L \times L$
simple cubic lattices.
According to Meirovitch
\cite{Meirovitch83},
we adopt the grand canonical ensemble.
The algorithm is described in the references
\cite{Meirovitch83,Yamagata95a}.

We start each simulation from a large activity and
then gradually decrease an activity.
The initial configurations have been obtained from preliminary
simulations.
The pseudorandom numbers are generated by the Tausworthe method
\cite{ItoKanada88,ItoKanada90}.
We measure physical quantities
over $10^{6}$ Monte Carlo steps per site
after discarding $5 \times 10^{4}$ Monte Carlo steps per site
to attain equilibrium.
We have checked that simulations from the ground state configuration
and no atom one gave consistent results.
Each run is divided into ten blocks.
Let us the average of a physical quantity, $O$, in each block
$\langle O \rangle_{i}$; $i = 1, 2, \ldots , 10$.
The expectation value is
\[
\overline{\langle O \rangle}
=
\frac{1}{10}\,\sum_{i=1}^{10} \langle O \rangle_{i}.
\]
The standard deviation is
\[
{\mit \Delta} \langle O \rangle
=
\left(
\overline{\langle O \rangle^{2}} - \overline{\langle O \rangle}^{2}
\right)^{1/2}/\sqrt{9}.
\]

Let us define a density by
\[
\rho
=
N/V
\]
where $N$ is the number of the atoms
in the lattice of $V$ sites
and an order parameter by
\[
R
=
2\,(N_{\rm A}-N_{\rm B})/V
\]
where $N_{\rm A}$ ($N_{\rm B}$) is the number of the atoms
in the A (B)-sublattice and $N = N_{\rm A} + N_{\rm B}$.
We measure the isothermal compressibility:
\begin{equation}
\kappa
=
V
\overline{
\left(
\langle \rho^{2} \rangle - \langle \rho \rangle^{2}
\right)
/ \langle \rho \rangle
},
\label{eqn:isocom}
\end{equation}
the staggered compressibility:
\begin{equation}
\chi^{\dagger}
=
V
\overline{
\left(
\langle R^{2} \rangle - \langle |R| \rangle^{2}
\right)
}
/ 4,
\label{eqn:stacom}
\end{equation}
and the fourth-order cumulant of $R$ \cite{Binder81}:
\begin{equation}
U
=
1-\frac{1}{3}\,
\overline{
\langle R^{4} \rangle / \langle R^{2} \rangle^{2}
}.
\label{eqn:cum}
\end{equation}

\section{Monte Carlo results}
\label{sec:mcr}
Figure 1 shows the activity dependence of
the isothermal compressibility, $\kappa_{L}(z)$,
defined by (\ref{eqn:isocom}) for various lattice sizes.
The solid curves are obtained by the smoothing procedure of
the {\em B\/}-spline
\cite{Tsuda88}.
As $L$ increases,
the shape of the curve becomes sharper.
There are shifts of the peak positions.
In figure~2  we show the activitiy dependence of
the staggered compressibility, $\chi_{L}^{\dagger}(z)$,
defined by (\ref{eqn:stacom}) for various lattice sizes.
It does not seem that the position of the peak shifts
to contrast those of $\kappa_{L}(z)$.
We show the activity dependence of the fourth-order cumulant,
$U_{L}(z)$, of $R$ defined by (\ref{eqn:cum})
for various lattice sizes in figure 3.
There is an intersection between the curves with
the size $L$ and $L+2$.
The positions of these intersections are within a narrow region.

We define effective critical activities,
$z_{\rm max}^{\kappa}(L)$ and $z_{\rm max}^{\chi^{\dagger}}(L)$,
as the peak position of $\kappa_{L}(z)$ and $\chi_{L}^{\dagger}(z)$,
respectively,
and $z_{\rm cross}^{U}(L)$ by
\begin{equation}
U_{L}(z_{\rm cross}^{U}(L))
=
U_{L+2}(z_{\rm cross}^{U}(L)).
\label{eqn:zcrossu}
\end{equation}
They will converge to the critical activity,
$z_{\rm c}$, as $L \to +\infty$.
We plot them against $1/L$ in figure 4.

We decide to estimate $z_{\rm c}$ from
$z_{\rm max}^{\chi^{\dagger}}(L)$ by the following reasons.
Although $z_{\rm max}^{\kappa}(L)$ seems to
behave systematically for $L \ge 6$,
it is difficult to extrapolate $z_{\rm c}$ from it
since we need a precise value of a critical exponent $\nu$:
$z_{\rm max}^{\kappa}(L) - z_{\rm c} \sim L^{-1/\nu}$.
We do not know the value of this system.
We cannot see systematic behaviour in
$z_{\rm max}^{\chi^{\dagger}}(L)$ and $z_{\rm cross}^{U}(L)$
for $L \ge 8$.
In the latter the value of $z_{\rm cross}^{U}(20)$ is deviate
from the others with $L \ge 12$.
We can see from the definition (\ref{eqn:zcrossu}) that it relates to
the values of $z_{\rm cross}^{U}(18)$ and $z_{\rm cross}^{U}(22)$.
Thus we do not adopt $z_{\rm cross}^{U}(L)$
as an estimator of $z_{\rm c}$
since we cannot ignore it simply.
We get the result, $z_{\rm c} = 0.7223(3)$,
by the arithmetic mean from the data $z_{\rm max}^{\chi^{\dagger}}(L)$
with $L$ = $12, 14, \ldots , 24$.
It consists with previous results,
$z_{\rm c} = 0.77(5)$
\cite{Gaunt67} and
$z_{\rm c} = 0.726(23)$
\cite{Racz80,Landau77},
within errors.
Our result is more precise than theirs.

\section{Summary}
\label{sec:sum}
We performed the Monte Carlo simulations of
the hard-sphere lattice gas with nearest neighbour exclusion
on the body-centred cubic lattice
under fully periodic boundary conditions.
We estimated the critical activity, $z_{\rm c}$,
to be $0.7223(3)$.
It consists with $z_{\rm c} = 0.77(5)$ by Gaunt
\cite{Gaunt67}.

As is described in section~\ref{sec:intro}
the critical activity relates to the slope of the critical curve
of the antiferromagnetic Ising model in the external field.
Our result, $a^{\ast}=0.1627(2)$, agrees
with $a^{\ast}=0.160(16)$ by Landau
\cite{Landau77}.
The system exhibits a re-entrant phase transition
since $a^{\ast}$ is positive.

In closing this paper we want to mention that
an antiferromagnetic Ising model in an external field is equivalent
to a Blume-Emery-Griffiths model~\cite{Griffiths67}.
Kasono and Ono~\cite{KasonoOno92} confirms that
there is a re-entrant phase transition
of the latter on the body-centred cubic lattice.
Our result supports theirs with high precision.

\section*{Acknowledgements}
The author would like to thank Dr. Katsumi Kasono
for useful discussions and critical reading of the manuscript.
We have carried out the simulations on the HITAC S-3600/120 computer
under the Institute of Statistical Mathematics Cooperative Research
Program (95-ISM$\cdot$CRP-37) and
on the personal computer with the Pentium/120MHz CPU and
the Linux 1.2.8 operating system
(Slackware-2.3.0 + JE-0.9.6-950717).
This study was supported by a Grant-in-Aid for Scientific Research
from the Ministry of Education, Science and Culture, Japan.


\clearpage
\section*{Figure captions}
\begin{description}
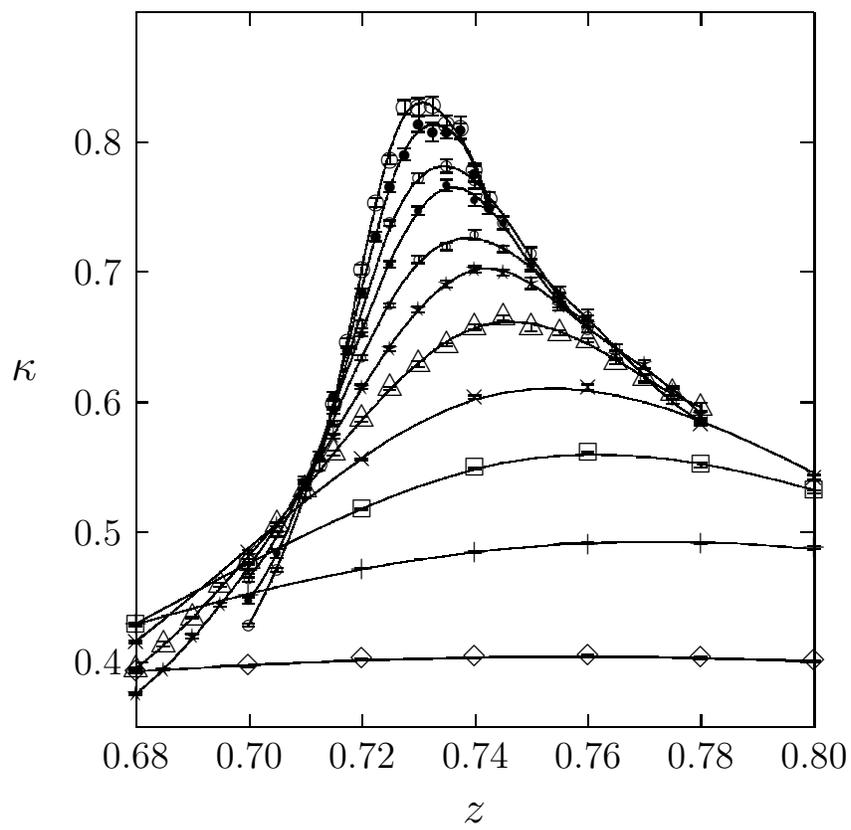
\item[Figure 1] Activity dependence of
the isothermal compressibility,
$\kappa$, defined by (\ref{eqn:isocom})
of the hard-sphere lattice gas (\ref{eqn:gpf})
on the body-centred cubic lattice
of $V$ sites under fully periodic boundary conditions.
$V = 2 \times L \times L \times L$;
$L$ = 4: $\Diamond$, 6: $+$, 8: $\Box$, 10: $\times$,
12: $\triangle$, 14: $\star$, 16: $\circ$(small),
18: $\bullet$(small), 20: $\circ$(middle), 22: $\bullet$(middle),
24: $\circ$(large).
The solid curves are obtained by the smoothing procedure
of the fourth-order {\em B\/}-spline.
The data with $L$ = 4, 6, $\ldots$ , 14 have been omitted to preserve
the clarity of the figure.

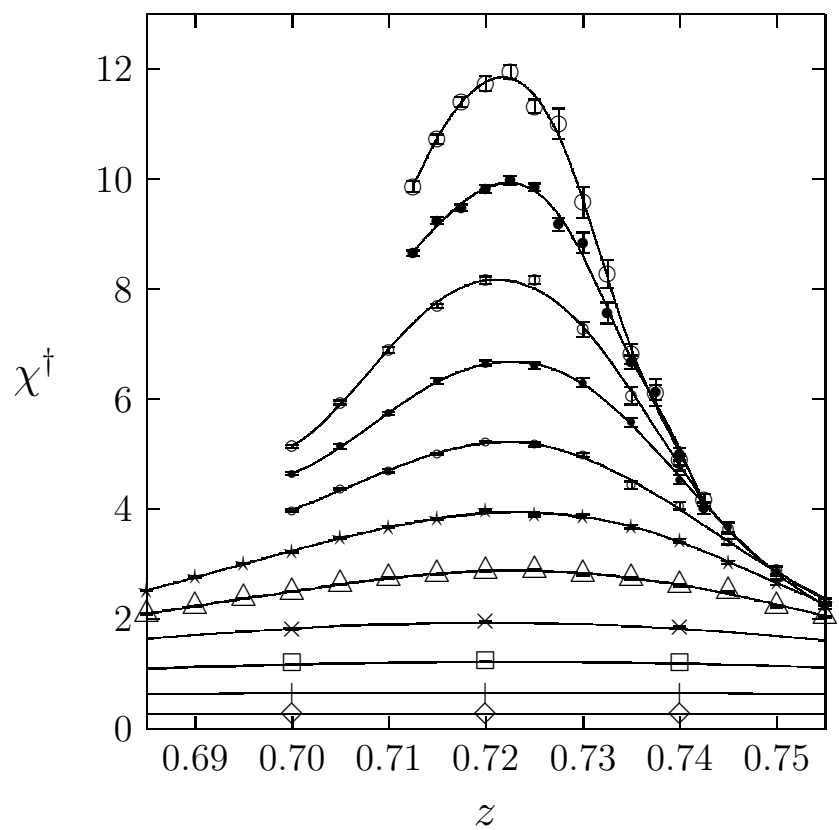
\item[Figure 2] Activity dependence of
the staggered compressibility,
$\chi^{\dagger}$, defined by (\ref{eqn:stacom}).
The meaning of the symbols and the curves is the same as in figure 1.
The data with $L$ = 4, 6, $\ldots$ , 20 have been omitted to preserve
the clarity of the figure.

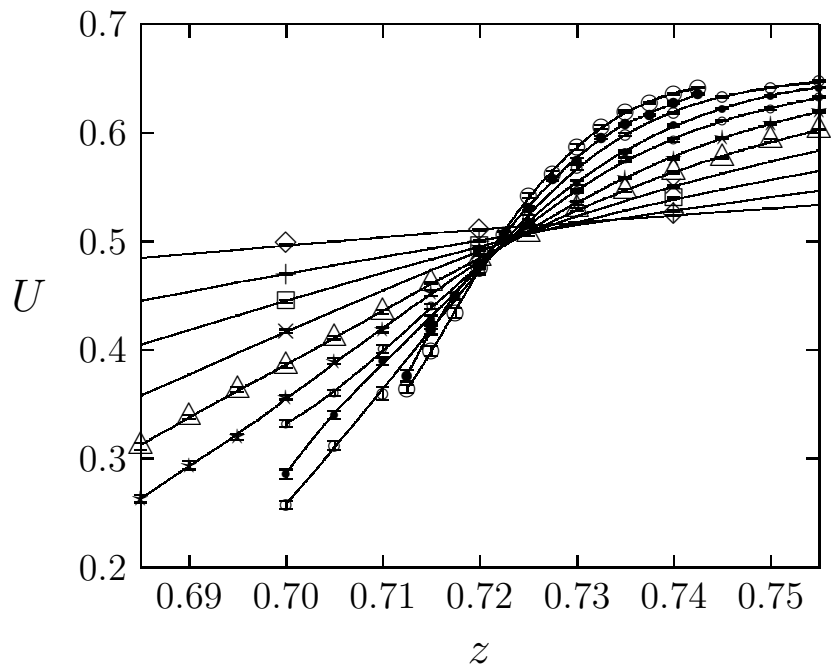
\item[Figure 3] Activitiy dependence of the fourth-order cumulant,
$U$, defined by~(\ref{eqn:cum}).
The meaning of the symbols and the curves is the same as in figure~1.
The data with $L$ = 4, 6, $\ldots$ , 20 have been omitted to preserve
the clarity of the figure.

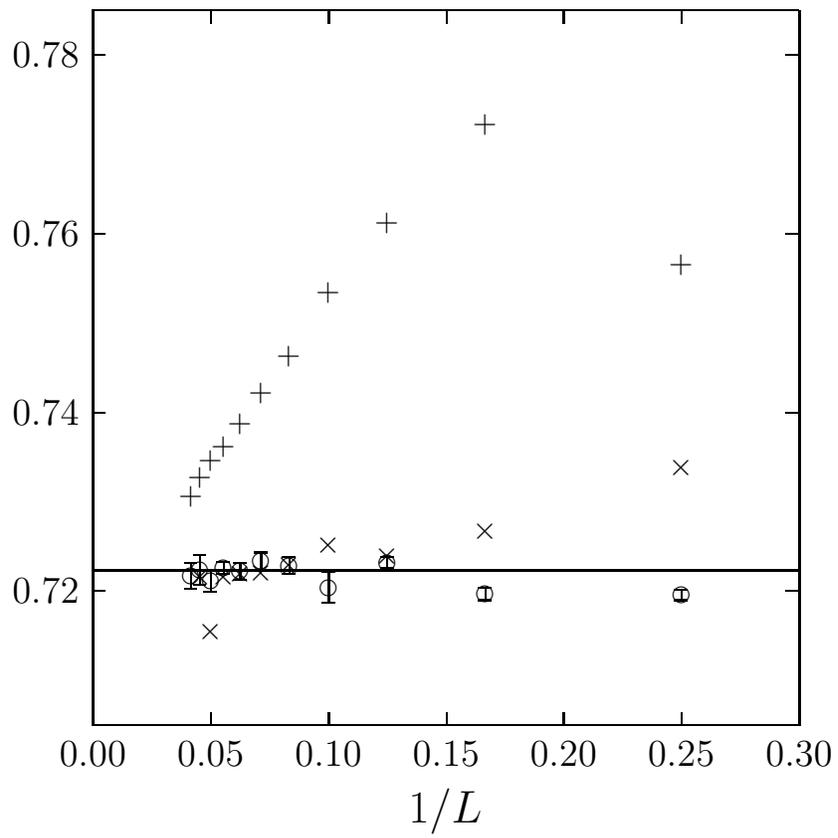
\item[Figure 4] Size dependence of
the effective critical activities,
$z_{\rm max}^{\kappa}(L)$: $+$,
$z_{\rm max}^{\chi^{\dagger}}(L)$: $\circ$,
and $z_{\rm cross}^{U}(L)$: $\times$.
Errors are less than the symbol size
for $z_{\rm max}^{\kappa}(L)$ and $z_{\rm cross}^{U}(L)$.
The horizontal line denotes $z = 0.7223$.
\end{description}

\clearpage
\pagestyle{empty}
\begin{figure}
\begin{center}
\setlength{\unitlength}{0.240900pt}
\ifx\plotpoint\undefined\newsavebox{\plotpoint}\fi
\sbox{\plotpoint}{\rule[-0.200pt]{0.400pt}{0.400pt}}%

\end{center}
\end{figure}

\clearpage
\begin{thebibliography}{99}
\bibitem{Domb58}
C. Domb,
Nuovo Ciment 9 Suppl. (1958) 9.

\bibitem{Temperley59}
H.N.V. Temperley,
Proc. Phys. Soc. 74 (1959) 183.

\bibitem{Burley60}
D.M. Burley,
Proc. Phys. Soc. 75 (1960) 262.

\bibitem{GauntFisher65}
D.S. Gaunt and M.E. Fisher,
J. Chem. Phys. 43 (1965) 2840.

\bibitem{Runnels65}
L.K. Runnels,
Phys. Rev. Lett. 15 (1965) 581.

\bibitem{RunnelsCombs66}
L.K. Runnels and L.L. Combs,
J. Chem. Phys. 45 (1966) 2482.

\bibitem{ReeChesnut66}
F.H. Ree and D..A. Chesnut,
J. Chem. Phys. 45 (1966) 3983.

\bibitem{Gaunt67}
D.S. Gaunt,
J. Chem. Phys. 46 (1967) 3237.

\bibitem{Baxteretal80}
R.J. Baxter, I.G. Enting, and S.K. Tsang,
J. Stat. Phys. 22 (1980) 465.

\bibitem{Baxter80}
R.J. Baxter,
J. Phys. A 13 (1980) L61.

\bibitem{Racz80}
Z. R\`acz,
Phys. Rev. B 21 (1980) 4012.

\bibitem{WoodGoldfinch80}
D.W. Wood and M. Goldfinch,
J. Phys. A 13 (1980) 2781.

\bibitem{BaxterPearce82}
R.J. Baxter and P.A. Pearce,
J. Phys. A 15 (1982) 897.

\bibitem{Meirovitch83}
H. Meirovitch,
J. Stat. Phys. 30 (1983) 681.

\bibitem{BloteWu90}
H.W.J. Bl\"ote and X.-N. Wu,
J. Phys. A 23 (1990) L627.

\bibitem{KamieniarzBlote93}
G. Kamieniarz and H.W.J. Bl\"ote,
J. Phys. A 26 (1993) 6679.

\bibitem{Yamagata95a}
A. Yamagata,
Physica A 215 (1995) 511.

\bibitem{Yamagata95b}
A. Yamagata,
Physica A in print.

\bibitem{Landau77}
D.P. Landau,
Phys. Rev. B 16 (1977) 4164.

\bibitem{Binder79}
K. Binder,
Monte Carlo Methods in Statistical Physics,
K. Binder, ed.
(Springer, Berlin, 1979) p. 1.

\bibitem{BinderStauffer87}
K. Binder and D. Stauffer,
Applications of the Monte Carlo Method in Statistical Physics,
2nd ed.,
K. Binder, ed.
(Springer, Berlin, 1987) p. 1.

\bibitem{ItoKanada88}
N. Ito and Y. Kanada,
Supercomputer 5 (1988) 31.

\bibitem{ItoKanada90}
N. Ito and Y. Kanada,
Proceedings of Supercomputing '90
(IEEE Computer Society Press, Los Alamitos, 1990) p. 753.

\bibitem{Binder81}
K. Binder,
Z. Phys. B 43 (1981) 119.

\bibitem{Tsuda88}
T. Tsuda,
Suchi Shori Programming
(Iwanami, Tokyo, 1988) p. 151.

\bibitem{Griffiths67}
R.B. Griffiths,
Physica 33 (1967) 689.

\bibitem{KasonoOno92}
K. Kasono and I. Ono,
Z. Phys. B 88 (1992) 213.
\end{thebibliography}
\end{document}